\documentclass[aps,prl,twocolumn,groupedaddress]{revtex4}
\usepackage{graphicx}% Include figure files
\usepackage{bm}% bold math
\newcommand{\etal}	{\textit{et~al.}}
\begin{document}

%\preprint{APS/123-QED}

\title{Quasiparticle cooling of a single-Cooper-pair-transistor}% Force line breaks with \\

\author{A. J. Ferguson}
\email{ajf1006@cam.ac.uk}%
\altaffiliation{%
Present address: Cavendish Laboratory, JJ Thomson Avenue, Cambridge, CB3 0HE, U.K.
}%
\affiliation{%
Australian Research Centre of Excellence for Quantum Computer
Technology, University of New South Wales, Sydney NSW 2052,
Australia
}%
\date{\today}% It is always \today, today,

\begin{abstract}
A superconducting tunnel junction is used to directly extract quasiparticles from one of the leads of a single-Cooper-pair-transistor. The consequent reduction in quasiparticle density causes a lower rate of quasiparticle tunneling onto the device. This rate is directly measured by radio-frequency reflectometry. Local cooling may be of direct benefit in reducing the effect of quasiparticles on coherent superconducting nanostructures.

\end{abstract}

\pacs{Valid PACS appear here}% PACS, the Physics and Astronomy
\maketitle

Coherent quantum nanostructures are highly sensitive to their thermal environment. In particular, superconducting single charge devices are strongly affected by heat in the form of quasiparticles. The quasiparticle has a pronounced effect on these devices due to its electronic charge \cite{joyez_prl_94}. If a quasiparticle tunnels from the leads onto the device island the electrostatic energy of the system changes, and Cooper pair coherence is destroyed \cite{lutchyn_prb_05, lutchyn_prb_07}. Due to the detrimental effect on Cooper-pair coherence this effect is often known as 'quasiparticle poisoning'. Recent experiments have used high-bandwidth techniques to determine quasiparticle tunneling rates \cite{aumentado_prl_04, ferguson_prl_06, shaw_preprint_08}.

The temperature of a nanostructure can be reduced by using on-chip, electronic refrigeration \cite{giazotto_rmp_06}. Superconductor-insulator-normal (SIN) tunnel junctions have been widely used for this purpose. This technique has been demonstrated to cool both metal islands \cite{manninen_apl_99,nahum_apl_94} and suspended dielectric membranes \cite{manninen_apl_97,clark_apl_05}. In this Letter a reduction in quasiparticle poisoning is demonstrated by extracting quasiparticles from the superconducting leads of a single-Cooper-pair-transistor (SCPT).

%%%%%%%%%%%%%%%%%%%%%%%%%%%%%%Figure1%%%%%%%%%%%%%%%%%%%%%%%%%%%%%%%%%%%%%%
%%%%%%%%%%%%%%%%%%%%%%%%%%%%%%%%%%%%%%%%%%%%%%%%%%%%%%%%%%%%%%%%%%%%%%%%%%%
\begin{figure}[h]
\begin{center}
\includegraphics[width=7.5cm]{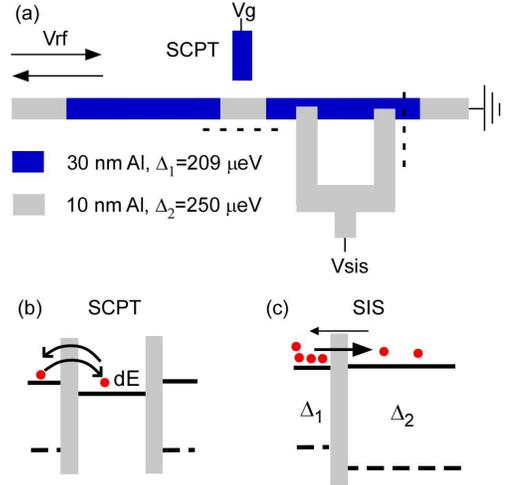}
\end{center}
\caption{(Color online) (a) Schematic of the SCPT and $S_1IS_2$ junction. The regions of two different gaps are indicated. Note that the lower gap region is isolated from the normal metal bond-pads (by the higher gap region) to avoid quasiparticle trapping. (b) Energy diagram of the SCPT, showing the potential minima for quasiparticles on the device island. (c) When the $S_1IS_2$ junction is biased to $eV_{S_1IS_2}=|\Delta_1-\Delta_2|$ quasiparticle extraction occurs from the SCPT reservoir.}\label{fig1trap}
\end{figure}
%%%%%%%%%%%%%%%%%%%%%%%%%%%%%%%%%%%%%%%%%%%%%%%%%%%%%%%%%%%%%%%%%%%%%%%%%%%

The nanostructure consists of a single-Cooper-pair-transistor (SCPT) and a S$_1$IS$_2$  junction (Fig. 1(a)). The SCPT, using the quasiparticle poisoning effect, enables relative measurements of quasiparticle density in its leads. The S$_1$IS$_2$ junction acts as a way to reduce the quasiparticle population. Quasiparticle density decreases exponentially with superconducting gap ($N_{qp}\propto\sqrt{T\Delta}\exp {-\frac{\Delta}{kT}}$). Therefore, when the singularities in the density-of-states are aligned by applying a bias ($V_{S_1IS_2}$) to the S$_1$IS$_2$ junction such that ($eV_{S_1IS_2}=\pm|\Delta_1-\Delta_2|$), the tunnel rate from the lower gap to the higher gap material is greater than in the reverse direction (Fig 1(c)) \cite{parmenter_prl_1961}. This allows cooling of the lower-gap region and the use of a S$_1$IS$_2$ junction as a refrigerator \cite{melton_prb_80}.

The device is fabricated by double angle evaporation of aluminium through a bilayer polymer resist mask. Between the evaporation stages a controlled oxidation is performed to define the tunnel barriers. Thin aluminium films, in which superconducting gap decreases with thickness \cite{meservery_71}, are used to generate the different gaps required for the S$_1$IS$_2$ junction. To achieve continuous thin films the sample is placed on a low temperature stage ($T\sim200$ K) during evaporation. The $\Delta_{1,2}$ regions have thicknesses of $30$ nm and  $10$ nm respectively. From previous measurements on SIS junctions it was found that $\Delta_{1}=209\pm11$ $\mu$eV and $\Delta_{2}=250\pm15$ $\mu$eV \cite{court_sst_07}. Device resistances, measured at 4.2 K, were $R_{SCPT}=17.7$ k$\Omega$ and $R_{S_1IS_2}=3.4$ k$\Omega$. The critical current of the S$_1$IS$_2$ junction was suppressed from a maximum of $I_c=69$ nA to $I_c<2$ nA by using a SQUID geometry.

Radio-frequency reflectometry is used to measure the SCPT \cite{schoelkopf_sci_1998}. The device is embedded in a LC circuit and placed at milliKelvin temperature in a dilution refrigerator. The circuit is resonant at 310 MHz and consists of a 470 nH chip inductor and a parasitic capacitance to ground. The complex reflection coefficient of a probe carrier signal at resonance depends on the presence of a quasiparticle on the SCPT. The carrier signal voltage biases the SCPT and its power is set to -94 dBm, chosen to maximize the signal to noise ratio. The reflected carrier signal is demodulated and time records are taken using an oscilloscope. Details of the apparatus have been previously published \cite{ferguson_prl_06}.

%%%%%%%%%%%%%%%%%%%%%%%%%%%%%%%%%%%%%%%%%%%%%%%%%%%%%%%%%%%%%%%%%%%%%%%%%%%%%%%%%%%%%%%%%%%%%%%%%%%%%%%FIGURE2%%%%%%%%%%%%%%%%%%%%%%%%%%%%%%%%%%%%%%%%
\begin{figure}[h]
\begin{center}
\includegraphics[width=8.0cm]{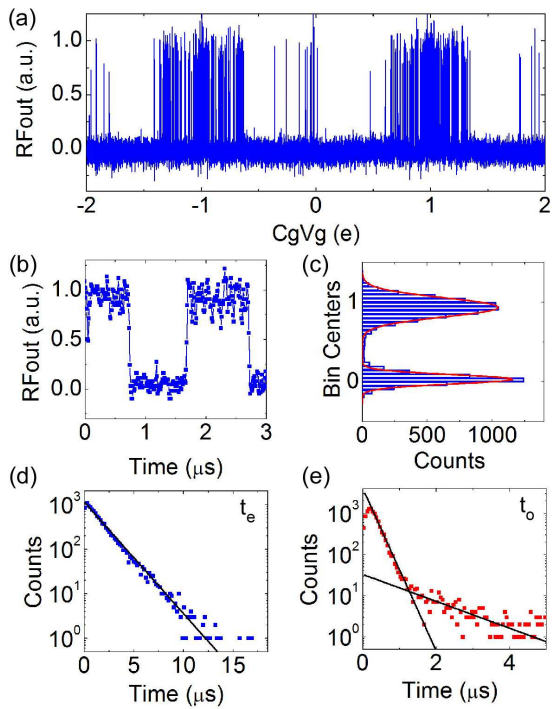}
\end{center}
\caption{(Color online) (a) Unaveraged gate sweep for the SCPT. (b) A single time trace at odd-integer charge. The points are separated by a time interval of 10 ns. (c) Histogram of the time-trace - solid lines are fits to a Gaussian distribution. (d) Example histogram showing distribution of even times (no quasiparticle on island). The solid line indicates a fit to an exponential giving a time constant of $t_e=1.7$ $\mu$s. (e) Example of an odd-time (quasiparticle on island) histogram showing dual-Poissonian distribution. The time constants are $t_{o1}= 22$ ns and $t_{o2}=1.3$ $\mu$s.}\label{fig2trap}
\end{figure}
%%%%%%%%%%%%%%%%%%%%%%%%%%%%%%%%%%%%%%%%%%%%%%%%%%%%%%%%%%%%%%%%%%%%%%%%%%%

In an unaveraged gate-sweep on the SCPT both the intrinsic behavior of the SCPT and the effect of quasiparticle tunneling are observed (Fig. 2(a)). In the absence of quasiparticles only the 2e periodic supercurrent oscillations would be present. With quasiparticles present in the leads, two level switching occurs on the supercurrent peaks as quasiparticles tunnel on and then off the island (Fig. 1(b)). A single shot time trace at the supercurrent maximum shows the bandwidth available with radio-frequency reflectometry (Fig. 2(b)). A rise time of approximately 30 ns is seen for a quasiparticle tunneling event. A histogram of the time trace gives a dual peak distribution (Fig. 2(c)) confirming that it is a two-level system. A real-time charge sensitivity is deduced, using $dQ=\frac{e}{SNR\times\sqrt{B}}$ \cite{cassidy_apl_07}. From the analysis described below the effective bandwidth of the measurement setup is determined to be 5 MHz. The charge sensitivity is $dQ=3.9\times 10^{-5}$ eHz$^{-0.5}$, a value similar to previous frequency domain measurements.

The time-traces are converted into a digital signal by comparing the recorded data with their median value. The times spent in the even (no quasiparticle) and odd (quasiparticle) states are then extracted and plotted in histograms. For the even times, the histogram is well-fitted by a single exponential (Fig. 2(d)) with time constant $t_e$. However, for the odd times, there are two time constants (Fig. 2(e)). This bi-exponential distribution is expected, due to the existence of both elastic and inelastic tunneling processes, and has been observed in recent measurements on Cooper pair boxes \cite{shaw_preprint_08}. The effect of finite measurement bandwidth is also apparent. It appears as the peak at short times in the histogram. The maxima occurs at 200 ns (Fig. 2(e)), indicating a system bandwidth of 5 MHz \cite{naaman_prl_06}.

%%%%%%%%%%%%%%%%%%%%%%%%%%%%%%%%%%%%%%%%%%%%%%%%%%%%%%%%%%%%%%%%%%%%%%%%%%%%%%%%%%%%%%%%%%%%%%%%%%%%%%%%%%Figure3%%%%%%%%%%%%%%%%%%%%%%%%%%%%%%%%%%%%%
\begin{figure}[h]
\begin{center}
\includegraphics[width=7.5cm]{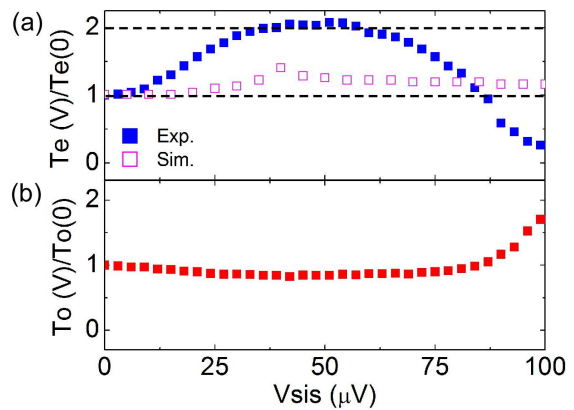}
\end{center}
\caption{(Color online) (a) The extracted even time constant $t_{e}$, normalised to $t_{e}(V_{ds}=0)=1.18$ $\mu s$, as a function of $S_1IS_2$ junction bias. There is a factor of two increase when $eV_{S_1IS_2}=\Delta_1-\Delta_2$ indicating cooling of one of the reservoirs. A simulation, based on balancing cooling power and heat transfer from the phonon system, is also shown. (b) The behavior of the odd time constant $t_{o2}$, normalised to $t_{o2}(V_{ds}=0)=58$ $ns$ under the same bias conditions.} \label{fig3trap}
\end{figure}
%%%%%%%%%%%%%%%%%%%%%%%%%%%%%%%%%%%%%%%%%%%%%%%%%%%%%%%%%%%%%%%%%%%%%%%%%%

The time constant $t_{e}$ is proportional to the quasiparticle density in the leads \cite{lutchyn_prb_05}. This allows relative changes in quasiparticle density to be determined. Due to the nature of the technique there is no, in principle, lower bound on the quasiparticle density that can be measured. At an experimental temperature of 250 mK the quasiparticle density in the leads is calculated to be $2.4\times10^{20}$ $m^{-3}$, leading to an average number of quasiparticles in the $\Delta_1$ regions of each lead of $\sim1$. The $t_{e}$ time constant is measured as a function of $S_1IS_2$ junction bias (Fig. 3(a)). It is seen that $t_{e}$ increases to a maximum at a voltage of $V_{S_1IS_2}=47$ $\mu$V. The maximum, expected on the basis of the aluminium film properties, $eV_{S_1IS_2}=\Delta_2-\Delta_1=41\pm18$ $\mu$eV, shows reasonable agreement with the experimental values. At the maximum $t_{e}$ is close to double its zero bias value indicating a two-fold reduction in quasiparticle density. There is only a two-fold reduction since quasiparticles are only being extracted from one of the leads. From the closeness to a factor of two, the contribution from the drain lead to quasiparticle tunneling is inferred to become negligible. Either quasi-holes or quasi-electrons are extracted, depending on the polarity of the $S_1IS_2$ junction bias. However the quasiparticle branches are strongly coupled, hence the effect on the time constants is symmetric with respect to bias polarity. The leads are separated from each other by the SCPT which regulates quasiparticle transfer by Coulomb blockade. As a result only a small effect of quasiparticle heat transfer through the SCPT is expected. Attempts to cool the source lead with an additional $S_1IS_2$ junction proved unsuccessful due to the radio-frequency signal being shorted out, this could be avoided by using a high value inductor (choke) on the source $S_1IS_2$ junction electrode to block the radio-frequency signal. This difficulty would also avoided in the case of a Cooper pair box which has only a single superconducting reservoir. The decrease in $t_e$ to past it's zero-bias value, at $V_{ds}=87$ $\mu$V, indicates an increase in quasiparticle density in the leads. This may be attributed to the onset of multi-particle tunneling processes \cite{adkins_rmp_64}.

A simulation of the change in $t_e$ as a function of bias was performed. The cooling power is provided by the $S_1IS_2$ junction and is balanced by heat transfer from the phonon system which remains close to the lattice temperature. The expression for the heat transfer from phonons to quasiparticles is given by $Q_{ph-qp}=\Sigma V (T_{ph}^5-T_{qp}^5)$ where $\Sigma=0.3\times10^9$ $Wm^{-3}K^{-5}$ is the electron-phonon coupling constant for aluminium and $V=2\times 0.1\times 0.03$ $\mu m^3$ is the volume of the $\Delta_1$ region of the drain electrode \cite{giazotto_rmp_06}. In addition, heat transfer ($Q_{S_1IS_2}$) through a $S_1IS_2$ junction can be numerically evaluated (Fig. 4(a)) \cite{frank_pla_97,manninen_apl_99}. By consistently solving $Q_{ph-qp}$ and $Q_{S_1IS_2}$ it is possible to determine a minimum quasiparticle temperature, and therefore a change in $t_e$. The result (Fig. 3(a)) significantly underestimates the observed effect. The assumption was made that the whole volume of the $\Delta_1$ region of the lead was cooled. However, one arm of the $S_1IS_2$ junction is in proximity ($\sim100$ $nm$) to the SCPT potentially leading to a much stronger quasiparticle extraction effect near the measuring device (the SCPT).

There is no significant change in $t_{o}$. As demonstrated in \cite{naaman_prl_06}, the effect of finite system bandwidth and the discrimination algorithm leads to an apparent change in one time constant as the other is varied. This can be corrected for but, since $t_e$ is the more interesting time constant in this experiment, the un-processed values of the time constants are presented.

%%%%%%%%%%%%%%%%%%%%%%%%%%%%%%%%%%%%%%%%%%%%%%%%%%%%%%%%%%%%%%%%%%%%%%%%%%%%%%%%%%%%%%%%%%%%%%%%%%%%%%%%%%Figure3%%%%%%%%%%%%%%%%%%%%%%%%%%%%%%%%%%%%%
\begin{figure}[h]
\begin{center}
\includegraphics[width=7.5cm]{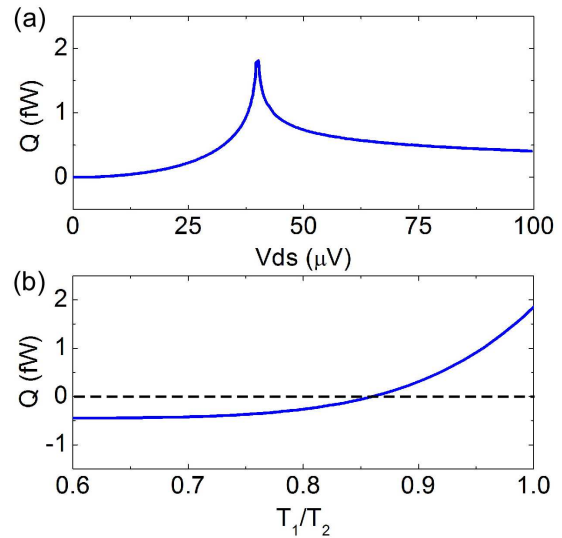}
\end{center}
\caption{(Color online) (a) Simulated cooling power ($Q_{S_1IS_2}$) of a $S_1IS_2$ junction, at constant temperature, versus bias. Calculation parameters are similar to the experimental case: T=250 mK, $\Delta_{1(2)}=210$ (250) $\mu$eV, $R=3.4$ k$\Omega$. (b) The cooling power of the same junction biased close to $eV_{S_1IS_2}=\Delta_2-\Delta_1$ as the temperature of the $\Delta_1$ material, corresponding to the lead of the SCPT, is reduced.} \label{fig3trap}
\end{figure}
%%%%%%%%%%%%%%%%%%%%%%%%%%%%%%%%%%%%%%%%%%%%%%%%%%%%%%%%%%%%%%%%%%%%%%%%%%

Finally, the cooling power is plotted (Fig. 4(a)) as a function of bias across the $S_1IS_2$ junction, showing the divergence at $eV_{S_1IS_2}=\pm|\Delta_1-\Delta_2|$, when the singularities in the density of states are aligned. A simulation was also performed of the cooling power, near it's maximum value at $V_{S_1IS_2}=40$ $\mu$V, as the temperature of $\Delta_1$ is reduced (Fig. 4(b)). The cooling power becomes negative for $\frac{T_1}{T_2}=0.86$, close to $\frac{\Delta_1}{\Delta_2}\sim\frac{T_1}{T_2}$, showing the maximum temperature reduction achievable. For an estimated initial temperature of 250 mK this would lead to a reduction in temperature to 215 mK and a consequent reduction in quasiparticle density, and increase in $t_e$, by a factor of 5. To achieve a larger temperature reduction, $\Delta_2$ could be increased by using thinner aluminium films or other superconducting materials. Alternatively, aluminium leads could be used with a lower gap material, such as titanium, for the island (therefore reducing $\Delta_1$) \cite{manninen_apl_99}.

In conclusion, this experiment demonstrates the potential of on-chip cooling to reduce the quasiparticle poisoning effect in superconducting coherent devices. Future experiments on Cooper pair boxes should be able to achieve a dramatic reduction in quasiparticle poisoning by means of quasiparticle extraction.

The author would like to thank R. G. Clark, N. Court, D. Barber and R. P. Starrett for supporting the experiments and P. Delsing for comments on the manuscript. This work is supported by the Australian Research Council, the Australian Government, and by the US National Security Agency (NSA) and US Army Research Office (ARO) under Contract No. W911NF-04-1-0290.

\bibliographystyle{apsrev}

\begin{thebibliography}{20}
\expandafter\ifx\csname natexlab\endcsname\relax\def\natexlab#1{#1}\fi
\expandafter\ifx\csname bibnamefont\endcsname\relax
  \def\bibnamefont#1{#1}\fi
\expandafter\ifx\csname bibfnamefont\endcsname\relax
  \def\bibfnamefont#1{#1}\fi
\expandafter\ifx\csname citenamefont\endcsname\relax
  \def\citenamefont#1{#1}\fi
\expandafter\ifx\csname url\endcsname\relax
  \def\url#1{\texttt{#1}}\fi
\expandafter\ifx\csname urlprefix\endcsname\relax\def\urlprefix{URL }\fi
\providecommand{\bibinfo}[2]{#2}
\providecommand{\eprint}[2][]{\url{#2}}

\bibitem[{\citenamefont{{P.~Joyez~\etal}}(1994)}]{joyez_prl_94}
\bibinfo{author}{\bibnamefont{{P.~Joyez}}},
\bibinfo{author}{\bibnamefont{{P.~Lafarge}}},
\bibinfo{author}{\bibnamefont{{A.~Filipe}}},
\bibinfo{author}{\bibnamefont{{D.~Esteve}}}, \bibnamefont{and}
\bibinfo{author}{\bibnamefont{{M.~Devoret}}},
\bibinfo{journal}{Phys. Rev.
  Lett.} \textbf{\bibinfo{volume}{72}}, \bibinfo{pages}{2458}
  (\bibinfo{year}{1994}).

\bibitem[{\citenamefont{Lutchyn et~al.}(2005)\citenamefont{Lutchyn, Glazman,
  and Larkin}}]{lutchyn_prb_05}
\bibinfo{author}{\bibfnamefont{R.}~\bibnamefont{Lutchyn}},
  \bibinfo{author}{\bibfnamefont{L.}~\bibnamefont{Glazman}}, \bibnamefont{and}
  \bibinfo{author}{\bibfnamefont{A.}~\bibnamefont{Larkin}},
  \bibinfo{journal}{Phys. Rev. B.} \textbf{\bibinfo{volume}{72}},
  \bibinfo{pages}{014517} (\bibinfo{year}{2005}).

\bibitem[{\citenamefont{Lutchyn and Glazman}(2007)}]{lutchyn_prb_07}
\bibinfo{author}{\bibfnamefont{R.~M.} \bibnamefont{Lutchyn}} \bibnamefont{and}
  \bibinfo{author}{\bibfnamefont{L.~I.} \bibnamefont{Glazman}},
  \bibinfo{journal}{Phys. Rev. B.} \textbf{\bibinfo{volume}{75}},
  \bibinfo{pages}{184520} (\bibinfo{year}{2007}).

\bibitem[{\citenamefont{Aumentado et~al.}(2004)\citenamefont{Aumentado, Keller,
  Martinis, and Devoret}}]{aumentado_prl_04}
\bibinfo{author}{\bibfnamefont{J.}~\bibnamefont{Aumentado}},
  \bibinfo{author}{\bibfnamefont{M.~W.} \bibnamefont{Keller}},
  \bibinfo{author}{\bibfnamefont{J.~M.} \bibnamefont{Martinis}},
  \bibnamefont{and} \bibinfo{author}{\bibfnamefont{M.~H.}
  \bibnamefont{Devoret}}, \bibinfo{journal}{Phys. Rev. Lett.}
  \textbf{\bibinfo{volume}{92}}, \bibinfo{pages}{066802}
  (\bibinfo{year}{2004}).

\bibitem[{\citenamefont{Ferguson et~al.}(2006)\citenamefont{Ferguson, Court,
  Hudson, and Clark}}]{ferguson_prl_06}
\bibinfo{author}{\bibfnamefont{A.~J.} \bibnamefont{Ferguson}},
  \bibinfo{author}{\bibfnamefont{N.}~\bibnamefont{Court}},
  \bibinfo{author}{\bibfnamefont{F.~E.} \bibnamefont{Hudson}},
  \bibnamefont{and} \bibinfo{author}{\bibfnamefont{R.~G.} \bibnamefont{Clark}},
  \bibinfo{journal}{Phys. Rev. Lett.} \textbf{\bibinfo{volume}{97}},
  \bibinfo{pages}{106603} (\bibinfo{year}{2006}).

\bibitem[{\citenamefont{Shaw et~al.}(2008)\citenamefont{Shaw, Lutchyn, Delsing,
  and Echternach}}]{shaw_preprint_08}
\bibinfo{author}{\bibfnamefont{M.~D.} \bibnamefont{Shaw}},
  \bibinfo{author}{\bibfnamefont{R.~M.} \bibnamefont{Lutchyn}},
  \bibinfo{author}{\bibfnamefont{P.}~\bibnamefont{Delsing}}, \bibnamefont{and}
  \bibinfo{author}{\bibfnamefont{P.~M.} \bibnamefont{Echternach}},
  \bibinfo{journal}{arXiv:0803.3102}  (\bibinfo{year}{2008}).

\bibitem[{\citenamefont{Giazotto et~al.}(2006)\citenamefont{Giazotto, Heikkilä,
  Luukanen, Savin, and Pekola}}]{giazotto_rmp_06}
\bibinfo{author}{\bibfnamefont{F.}~\bibnamefont{Giazotto}},
  \bibinfo{author}{\bibfnamefont{T.~T.} \bibnamefont{Heikkilä}},
  \bibinfo{author}{\bibfnamefont{A.}~\bibnamefont{Luukanen}},
  \bibinfo{author}{\bibfnamefont{A.~M.} \bibnamefont{Savin}}, \bibnamefont{and}
  \bibinfo{author}{\bibfnamefont{J.~P.} \bibnamefont{Pekola}},
  \bibinfo{journal}{Rev. Mod. Phys} \textbf{\bibinfo{volume}{78}},
  \bibinfo{pages}{217} (\bibinfo{year}{2006}).

\bibitem[{\citenamefont{Manninen et~al.}(1999)\citenamefont{Manninen,
  Suoknuuti, Leivo, and Pekkola}}]{manninen_apl_99}
\bibinfo{author}{\bibfnamefont{A.~J.} \bibnamefont{Manninen}},
  \bibinfo{author}{\bibfnamefont{J.~K.} \bibnamefont{Suoknuuti}},
  \bibinfo{author}{\bibfnamefont{M.~M.} \bibnamefont{Leivo}}, \bibnamefont{and}
  \bibinfo{author}{\bibfnamefont{J.~P.} \bibnamefont{Pekkola}},
  \bibinfo{journal}{Appl. Phys. Lett.} \textbf{\bibinfo{volume}{74}},
  \bibinfo{pages}{3020} (\bibinfo{year}{1999}).

\bibitem[{\citenamefont{Nahum et~al.}(1994)\citenamefont{Nahum, Eiles, and
  Martinis}}]{nahum_apl_94}
\bibinfo{author}{\bibfnamefont{M.}~\bibnamefont{Nahum}},
  \bibinfo{author}{\bibfnamefont{T.~M.} \bibnamefont{Eiles}}, \bibnamefont{and}
  \bibinfo{author}{\bibfnamefont{J.~M.} \bibnamefont{Martinis}},
  \bibinfo{journal}{Appl. Phys. Lett.} \textbf{\bibinfo{volume}{65}},
  \bibinfo{pages}{3123} (\bibinfo{year}{1994}).

\bibitem[{\citenamefont{Manninen et~al.}(1997)\citenamefont{Manninen, Leivo,
  and Pekkola}}]{manninen_apl_97}
\bibinfo{author}{\bibfnamefont{A.~J.} \bibnamefont{Manninen}},
  \bibinfo{author}{\bibfnamefont{M.~M.} \bibnamefont{Leivo}}, \bibnamefont{and}
  \bibinfo{author}{\bibfnamefont{J.~P.} \bibnamefont{Pekkola}},
  \bibinfo{journal}{Appl. Phys. Lett.} \textbf{\bibinfo{volume}{70}},
  \bibinfo{pages}{1885} (\bibinfo{year}{1997}).

\bibitem[{\citenamefont{Clark et~al.}(2005)\citenamefont{Clark, Miller,
  Williams, Ruggiero, Hilton, Vale, Beall, Irwin, and Ullom}}]{clark_apl_05}
\bibinfo{author}{\bibfnamefont{A.~M.} \bibnamefont{Clark}},
  \bibinfo{author}{\bibfnamefont{N.~A.} \bibnamefont{Miller}},
  \bibinfo{author}{\bibfnamefont{A.}~\bibnamefont{Williams}},
  \bibinfo{author}{\bibfnamefont{S.~T.} \bibnamefont{Ruggiero}},
  \bibinfo{author}{\bibfnamefont{G.~C.} \bibnamefont{Hilton}},
  \bibinfo{author}{\bibfnamefont{L.~R.} \bibnamefont{Vale}},
  \bibinfo{author}{\bibfnamefont{J.~A.} \bibnamefont{Beall}},
  \bibinfo{author}{\bibfnamefont{K.~D.} \bibnamefont{Irwin}}, \bibnamefont{and}
  \bibinfo{author}{\bibfnamefont{J.~N.} \bibnamefont{Ullom}},
  \bibinfo{journal}{Appl. Phys. Lett.} \textbf{\bibinfo{volume}{86}},
  \bibinfo{pages}{173508} (\bibinfo{year}{2005}).

\bibitem[{\citenamefont{Parmenter}(1961)}]{parmenter_prl_1961}
\bibinfo{author}{\bibfnamefont{R.~H.} \bibnamefont{Parmenter}},
  \bibinfo{journal}{Phys. Rev. Lett.} \textbf{\bibinfo{volume}{7}},
  \bibinfo{pages}{274} (\bibinfo{year}{1961}).

\bibitem[{\citenamefont{Melton et~al.}(1980)\citenamefont{Melton, Paterson, and
  Kaplan}}]{melton_prb_80}
\bibinfo{author}{\bibfnamefont{R.~G.} \bibnamefont{Melton}},
  \bibinfo{author}{\bibfnamefont{J.~L.} \bibnamefont{Paterson}},
  \bibnamefont{and} \bibinfo{author}{\bibfnamefont{S.~B.}
  \bibnamefont{Kaplan}}, \bibinfo{journal}{Phys. Rev. B.}
  \textbf{\bibinfo{volume}{21}}, \bibinfo{pages}{1858} (\bibinfo{year}{1980}).

\bibitem[{\citenamefont{Meservey and Tedrow}(1971)}]{meservery_71}
\bibinfo{author}{\bibfnamefont{R.}~\bibnamefont{Meservey}} \bibnamefont{and}
  \bibinfo{author}{\bibfnamefont{P.~M.} \bibnamefont{Tedrow}},
  \bibinfo{journal}{J. Appl. Phys.} \textbf{\bibinfo{volume}{42}},
  \bibinfo{pages}{51} (\bibinfo{year}{1971}).

\bibitem[{\citenamefont{Court et~al.}(2007)\citenamefont{Court, Ferguson, and
  Clark}}]{court_sst_07}
\bibinfo{author}{\bibfnamefont{N.~A.} \bibnamefont{Court}},
  \bibinfo{author}{\bibfnamefont{A.~J.} \bibnamefont{Ferguson}},
  \bibnamefont{and} \bibinfo{author}{\bibfnamefont{R.~G.} \bibnamefont{Clark}},
  \bibinfo{journal}{Supercond. Sci. Technol.} \textbf{\bibinfo{volume}{21}},
  \bibinfo{pages}{015013} (\bibinfo{year}{2007}).

\bibitem[{\citenamefont{Schoelkopf et~al.}(1998)\citenamefont{Schoelkopf,
  Wahlgren, Kozhevnikov, Delsing, and Prober}}]{schoelkopf_sci_1998}
\bibinfo{author}{\bibfnamefont{R.~J.} \bibnamefont{Schoelkopf}},
  \bibinfo{author}{\bibfnamefont{P.}~\bibnamefont{Wahlgren}},
  \bibinfo{author}{\bibfnamefont{A.~A.} \bibnamefont{Kozhevnikov}},
  \bibinfo{author}{\bibfnamefont{P.}~\bibnamefont{Delsing}}, \bibnamefont{and}
  \bibinfo{author}{\bibfnamefont{D.~E.} \bibnamefont{Prober}},
  \bibinfo{journal}{Science} \textbf{\bibinfo{volume}{280}},
  \bibinfo{pages}{1238} (\bibinfo{year}{1998}).

\bibitem[{\citenamefont{Cassidy et~al.}(2007)\citenamefont{Cassidy, Dzurak,
  Clark, Petersson, Farrer, Ritchie, and Smith}}]{cassidy_apl_07}
\bibinfo{author}{\bibfnamefont{M.~C.} \bibnamefont{Cassidy}},
  \bibinfo{author}{\bibfnamefont{A.~S.} \bibnamefont{Dzurak}},
  \bibinfo{author}{\bibfnamefont{R.~G.} \bibnamefont{Clark}},
  \bibinfo{author}{\bibfnamefont{K.~D.} \bibnamefont{Petersson}},
  \bibinfo{author}{\bibfnamefont{I.}~\bibnamefont{Farrer}},
  \bibinfo{author}{\bibfnamefont{D.~A.} \bibnamefont{Ritchie}},
  \bibnamefont{and} \bibinfo{author}{\bibfnamefont{C.~G.} \bibnamefont{Smith}},
  \bibinfo{journal}{Appl. Phys. Lett.} \textbf{\bibinfo{volume}{91}},
  \bibinfo{pages}{222104} (\bibinfo{year}{2007}).

\bibitem[{\citenamefont{Naaman and Aumentado}(2006)}]{naaman_prl_06}
\bibinfo{author}{\bibfnamefont{O.}~\bibnamefont{Naaman}} \bibnamefont{and}
  \bibinfo{author}{\bibfnamefont{J.}~\bibnamefont{Aumentado}},
  \bibinfo{journal}{Phys. Rev. Lett.} \textbf{\bibinfo{volume}{96}},
  \bibinfo{pages}{100201} (\bibinfo{year}{2006}).

\bibitem[{\citenamefont{Adkins}(1964)}]{adkins_rmp_64}
\bibinfo{author}{\bibfnamefont{C.~J.} \bibnamefont{Adkins}},
  \bibinfo{journal}{Rev. Mod. Phys.} \textbf{\bibinfo{volume}{36}},
  \bibinfo{pages}{212} (\bibinfo{year}{1964}).

\bibitem[{\citenamefont{Frank and Kresh}(1997)}]{frank_pla_97}
\bibinfo{author}{\bibfnamefont{B.}~\bibnamefont{Frank}} \bibnamefont{and}
  \bibinfo{author}{\bibfnamefont{W.}~\bibnamefont{Kresh}},
  \bibinfo{journal}{Phys. Lett. A} \textbf{\bibinfo{volume}{235}},
  \bibinfo{pages}{281} (\bibinfo{year}{1997}).

\end{thebibliography}

\end{document}